\documentclass[usenatbib]{mn2e}
\usepackage{graphicx}
\usepackage{amssymb}
\usepackage{amsmath}

\title[Timescales in Sheets and Filaments]{Accretion and Diffusion Timescales in Sheets and Filaments}
\author[F. Heitsch and L. Hartmann]{F. Heitsch$^{1}$\thanks{E-mail:
fheitsch@unc.edu (FH); lhartm@umich.edu (LH)} and L. Hartmann$^{2}$\\
$^{1}$Department of Physics and Astronomy, University of North Carolina Chapel Hill, Chapel Hill, NC 27599-3255, U.S.A\\
$^{2}$Department of Astronomy, University of Michigan, Ann Arbor, MI 48109, U.S.A}

\begin{document}

\date{Accepted 2014 June 08.  Received 2014 June 08; in original form 2014 March 05}

\pagerange{\pageref{firstpage}--\pageref{lastpage}} \pubyear{---}

\maketitle

\label{firstpage}

\begin{abstract}
A comparison of accretion and (turbulent) magnetic
diffusion timescales for sheets and filaments demonstrates that 
dense star-forming clouds generally will -- under realistic 
conditions -- become supercritical due to mass accretion on timescales at least an order of magnitude shorter than 
ambipolar and/or turbulent diffusion timescales. Thus, ambipolar or turbulent diffusion -- while present -- is unlikely to control the formation of cores
and stars. 
\end{abstract}


\section{Introduction}\label{s:introduction}

The role of magnetic fields in star formation has been debated for many decades.  
\citet{1956MNRAS.116..503M} showed that spherical or isotropic contraction from conditions in the
diffuse interestellar medium would prevent gravitationally-bound solar-mass protostellar clouds
from forming unless substantial amounts of magnetic flux were lost during contraction.  They 
suggested that this flux loss was a result of ambipolar diffusion in dark clouds 
where the ion fraction becomes very low due to dust shielding of the interstellar
radiation field.  Following this work, a large number of investigations have been dedicated over
the years to examining the effects of magnetic fields on the rotation and structure of protostellar
clouds and their time evolution 
\citep[][see also \citealp{1987ARA&A..25...23S}]{1979ApJ...228..475M,1979ApJ...228..159M,1993ApJ...418..774C,2000ApJ...529..925C,2006ApJ...652..442C}.

In its most extreme form, the picture of protostellar cloud contraction controlled by ambipolar diffusion
led to the concept of ``slow'' star formation \citep{1987ARA&A..25...23S}.  The lengthy diffusion timescales
in principle would then explain the low efficiencies of star formation in
molecular clouds \citep{1979ApJ...227L.105C,2009ApJS..181..321E}.  However, studies of the stellar populations
in molecular clouds show that the stars are generally young - a few Myr old at most -, and they show no
evidence for substantial numbers of older ($\sim 10$~Myr-old) stars.  This has led to the countervailing
picture of ``rapid'' star formation \citep{2001ApJ...562..852H}, in which
ambipolar diffusion does not play a major role in lengthening collapse timescales.  In this picture
the efficiency of star formation is not determined by slow contraction but rather by 
cloud dispersal via stellar energy input 
\citep{1994ApJ...436..795F,2005MNRAS.358..291D,2012MNRAS.427..625W}\footnote{\citet{2012MNRAS.427..625W}
point out that the role of stellar feedback depends on the cloud mass \citep{2012MNRAS.424..377D}. Also,
once the cloud is allowed to continue to accrete material during the feedback phase, feedback
not necessarily disperses the cloud, but keeps gas from reaching the high densities necessary for star formation 
\citep{2010ApJ...715.1302V}.}, or possibly by tidal forces for lower-mass clouds 
\citep{2009MNRAS.393.1563B,2009MNRAS.395L..81B}.

A mechanism which might reduce the timescale for the diffusion of magnetic flux is turbulence.
This possibility has been explored
in the context of turbulent ambipolar diffusion 
\citep{2002ApJ...567..962Z,2002ApJ...578L.113K,2002ApJ...570..210F,2004ApJ...603..165H,2004ApJ...609L..83L,2008ApJ...679L..97K,2011ApJ...728..123K},
and of reconnection diffusion \citep{1999ApJ...517..700L,2010ApJ...714..442S,2011ApJ...743...51E,2013Natur.497..466E}.

Alternatively, \citet{1956MNRAS.116..503M} recognized that the ``magnetic flux problem'' could be avoided
by mass accumulation along field lines, which could achieve the necessary mass-to-magnetic flux ratio
without reducing the magnetic flux.  They rejected this solution, however, by arguing
that the mass would have to be accumulated over such long length scales that it would gravitationally
fragment.  We now understand that this argument is not applicable because the interstellar medium is
generally in supersonic motion, and the usual Jeans criterion based on balancing thermal pressure against
gravity is irrelevant.  Indeed, numerical simulations of the large-scale galactic interstellar medium show
precisely this accumulation of mass along field lines 
\citep[][see also \citealp{2006ApJ...646..213K}]{1995ApJ...441..702V,2001ApJ...562..852H}.

Support for the idea of mass accumulation along field lines comes from
the comprehensive observational analysis of \citet{2010ApJ...725..466C}, who found
no evidence for increasing average field strength with increasing density below $n\sim 300$~cm$^{-3}$ 
\citep[see also][but \citealp{2012ApJ...755..130M} for newer results from polarimetry]{1986ApJ...301..339T}. 
At higher densities, the maximum field strength
increases as $B \propto n^{2/3}$, which Crutcher et al. suggest is due to isotropic contraction under gravity.
In addition, recent studies of star formation occurring in filamentary gas structures, which appears to
be common \citep{2010A&A...518L.102A}, also provide support for the idea of building dense structures
initially via flows along the magnetic field.  In some well-studied cases, such as B216/17, B18, L1506 
\citep{1987ApJ...321..855H,1990ApJ...359..363G,1984ApJ...282..508M} and
B211/13 in Taurus \citep{2013A&A...550A..38P}, and the Pipe Nebula \citep{2008A&A...486L..13A}
the magnetic field appears relatively ordered and nearly perpendicular to the filamentary gas, as
would be expected in this scenario.  Furthermore, \citet{2013A&A...550A..38P}
show that the CO emission around B211 exhibits differing radial velocities on either
side of the filament, interpreting this as evidence for inflow of ambient molecular gas.
For a systematic study of the alignment of ambient magnetic fields with local molecular clouds,
see \citet{2013MNRAS.436.3707L}.

These observational results motivate us to
address the question of ambipolar and turbulent diffusion vs. mass accumulation using a relatively
simple geometric setup. We compare the accretion and diffusion timescales for 
magnetized sheets (Sec.~\ref{s:sheet})
and filaments (Sec.~\ref{s:filament}), based on an earlier study discussing the evolution 
of the Pipe nebula \citep{2009ApJ...704.1735H}. We find for both sheet- and filament-like geometries
that accretion will dominate a cloud's evolution from subcritical to supercritical. 

\section{Accretion and Ambipolar/Turbulent Diffusion Timescales for a Sheet}\label{s:sheet}
We assume that 
star-forming molecular clouds are assembled by flows along magnetic field lines. 
This is reasonable given the fact (a) that perpendicular field components can efficiently prevent
compression and high density contrasts \citep{2008ApJ...687..303I,2009ApJ...704..161I,2009ApJ...695..248H}, and (b) that 
magnetic field vectors seem aligned perpendicularly to filamentary structures 
\citep[][see also \citet{2002ApJ...578..914H} for a scenario]{1987ApJ...321..855H,2013A&A...550A..38P}.
We envision cloud formation to occur conceptually in two phases.  The first involves the accumulation of
lower-density gas as a result of large-scale flows in the interstellar medium (Sec.~\ref{ss:sweepupsheet}).  
This has usually been considered
in terms of the sweep-up of diffuse atomic gas \citep{1995ApJ...441..702V,2001ApJ...562..852H}, though there is
no reason why, in environments other than the solar neighborhood, accumulation of partially or mostly diffuse
molecular gas cannot occur as well \citep{2013ApJ...779...43P}.
The second phase addresses the observation that star formation occurs in the densest regions of molecular clouds,
where such regions generally represent a small fraction of the total cloud mass
\citep{2010ApJ...724..687L, 2010ApJ...723.1019H}.
We explore this later phase (Sec.~\ref{ss:gravitysheet}) by considering the development of denser sheets and filaments within the molecular cloud
driven by self-gravity \citep[e.g.][]{2007ApJ...657..870V}.

\subsection{Cloud Formation by Sweep-up of Diffuse Gas}\label{ss:sweepupsheet}

We assume that the cloud is forming due to the collision of two identical flows of diffuse atomic gas, with 
constant density $n_0$ and velocity $v_0$. The flows are parallel to the background magnetic field $B$. Then, the 
cloud can be treated as a "sheet" of column density $N_H$,
accreting gas at a rate $2n_0 v_0$, or
\begin{eqnarray}
  N_H&=& 2n_0 v_0 t\label{e:Nacc}\\
     &=& 6.31\times10^{20}\left(\frac{n_0}{\mbox{cm}^{-3}}\right)
   \left(\frac{v_0}{10 \mbox{km s}^{-1}}\right)\left(\frac{t}{\mbox{Myr}}\right)\mbox{ cm}^{-2}.\nonumber
\end{eqnarray}
A thin sheet that is permeated by a magnetic field $B$ can fragment if 
\begin{equation}
  N_H > \frac{B}{2\pi m G^{1/2}},\label{e:Ncrit}
\end{equation}
or, if the mass-to-flux ratio at a given mass column density $\Sigma=mN_H$ is larger than the critical value,
\begin{equation}
  \frac{\Sigma}{B}>\left(\frac{\Sigma}{B}\right)_c \equiv \frac{1}{2\pi G^{1/2}},
\end{equation}
\citep{1978PASJ...30..671N}. The time $\tau_{acc}$ needed for the cloud to reach criticality results 
from equations~\ref{e:Nacc} and \ref{e:Ncrit},
\begin{eqnarray}
  \tau_{acc}&=& \frac{B}{4\pi n_0 v_0 m G^{1/2}}\label{e:tauc}\\
            &=& 5.65\left(\frac{B}{5 \mu\mbox{G}}\right)
  \left(\frac{n_0}{\mbox{ cm}^{-3}}\right)^{-1}\left(\frac{v_0}{10 \mbox{km s}^{-1}}\right)^{-1} \mbox{ Myr}.\nonumber
\end{eqnarray}

The ambipolar diffusion timescale at a given length scale $L$ can be written as (see App.~\ref{a:adtime})
\begin{equation}
  \tau_{AD} = 1.38\times 10^3 \left(\frac{L}{\mbox{pc}}\right)^2\left(\frac{n}{3\times 10^2\mbox{cm}^{-3}}\right)^{3/2}\left(\frac{B}{5 \mu\mbox{G}}\right)^{-2}\mbox{ Myr}.
  \label{e:tAD}
\end{equation}
Note that we use a different scaling for the density corresponding to the transition from $B\propto n^0$ to $B\propto n^{2/3}$ \citep{2010ApJ...725..466C}, 
since we are interested in ambipolar diffusion {\em within} the sheet-like cloud. 
Comparing equations~\ref{e:tauc} and \ref{e:tAD} suggests that accretion 
may be faster than ambipolar diffusion on the whole. 

More specifically, we discuss two regimes. (1) For a cloud forming from the diffuse interstellar gas via sweep-up, we set the flow densities and
velocities to $1$~cm$^{-3}$ and $10$~km~s$^{-1}$. For the (still) diffuse, forming cloud, we assume $n=300$~cm$^{-3}$, and a length scale of a parsec.
For these numbers, accretion is faster than ambipolar diffusion ($\tau_{acc}<\tau_{AD}$) if $B\lesssim 30\mu$G. Note that ambipolar diffusion will not be relevant 
before a (UV-shielding) column density corresponding to $A_V=1$ (or $B\approx 8\mu$G) has been accumulated (see App.~\ref{a:adtime}).
(2) Once sufficient mass has been accumulated, the accretion flows will start to be driven by gravity (see Sec.~\ref{ss:gravitysheet}). In this case, 
we assume a flow density of $100$~cm$^{-3}$, and a flow velocity of $1$~km~s$^{-1}$. At the same lengthscale of $1$~pc, $\tau_{acc}<\tau_{AD}$ for
$B\lesssim 540\mu$G. At a tenth of a parsec, the condition is reached for $B\lesssim 5.4\mu$G. 

\subsection{Dense Gas Formation by Gravitationally Driven Flows}\label{ss:gravitysheet}

We next follow the evolution of a sheet-like cloud that is allowed to accrete gas at free-fall velocities.
Here we are thinking of the accumulation of diffuse molecular gas - usually constituting the majority of
the cloud - 
driven by gravity
into a dense structure that can form stars, adopting a sheet geometry for simplicity
(we consider filaments in Sec.~\ref{s:filament}).
The gravitational acceleration towards an infinite sheet 
with mass column density $\Sigma$ is given by 
\begin{equation}
  \left|a_s\right| = 2\pi G\Sigma\,.
\end{equation}
For steady-state accretion, the velocity profile can be written as
\begin{equation}
  v_z = -2(\pi G\Sigma(z_{ref}-z))^{1/2},
  \label{e:vzsheet}
\end{equation}
{\em if} we assume that $\Sigma$ does not vary strongly while the column of ambient gas
\begin{equation}
  \Sigma_{acc}\equiv2\rho_0 z_{ref} 
\end{equation}
is accreted. We will justify this assumption in Appendix~\ref{a:nocolumn}. Here, 
$z_{ref}$ is a reference distance to be chosen, and will be on the order of $1$~pc.  
For a field strength of $10\mu$G (using eq.~\ref{e:Ncrit}), this results in an infall velocity
of $1$~km~s$^{-1}$, consistent with observational estimates 
\citep{2013ApJ...766..115K,2013MNRAS.436.1513F,2013A&A...550A..38P}.

We ignore the ram pressure of the infalling gas 
(see App.~\ref{a:lowpress}). Thus, we can set $z$ in equation~\ref{e:vzsheet} to
the scaleheight of an isothermal sheet (at given sound speed $c_s$),
\begin{equation}
  H = \frac{c_s^2}{\pi G \Sigma}.
  \label{e:scaleheight}
\end{equation}
Then, the accretion rate onto the sheet is given by
\begin{eqnarray}
  \dot{\Sigma} &=& 2\rho_0 v_z\nonumber\\
               &=& 4\rho_0\left(\pi G z_{ref} \Sigma\left(1-\frac{H}{z_{ref}}\right)\right)^{1/2}\nonumber\\
               &=& a\left(\Sigma-b\right)^{1/2}.\label{e:sigmadotsheet}
\end{eqnarray}
In the last step we defined 
\begin{eqnarray}
  a&\equiv&4\rho_0\left(\pi G z_{ref}\right)^{1/2}\label{e:apar}\\
  b&\equiv&\frac{H}{z_{ref}}\Sigma=\frac{c_s^2}{\pi G z_{ref}}\label{e:bpar}.
\end{eqnarray}
The solution to eq.~\ref{e:sigmadotsheet} is 
\begin{equation}
  \Sigma(t) = \frac{1}{4}\left(a^2 t^2+2a\Sigma_0^{1/2}t+4b+\Sigma_0\right),
  \label{e:sigmat}
\end{equation}
with the initial column density $\Sigma_0 = m N(t=0)$ to be chosen.
Equating this with eq.~\ref{e:Ncrit} and solving for the accretion time $\tau_{acc}$ yields 
\begin{equation}
  \tau_{acc} = \frac{1}{a}\left(2\left(\frac{B}{2\pi G^{1/2}}-b\right)^{1/2}-\Sigma_0^{1/2}\right).
  \label{e:taccsheet}
\end{equation}
To develop physical insight, we further simplify equation~\ref{e:taccsheet}. First, we can choose
$\Sigma_0$, the initial column density, to be negligibly small, i.e. $\Sigma_0\ll \Sigma_c\equiv B/(2\pi\sqrt{G})$. 
Second, $b$ is essentially the ratio of $H/z_{ref}$, which for an evolved sheet will also be small, if we set $z_{ref}=3$~pc.
With these simplifications, equation~\ref{e:taccsheet} turns into
\begin{equation}
  \tau_{acc}\approx (2\pi G \rho_0)^{-1/2} \left(\frac{\Sigma_c}{\Sigma_{acc}}\right)^{1/2},
  \label{e:taccsheetsimple}
\end{equation}
i.e. the timescale to reach criticality is given by the free-fall time of the ambient gas multiplied by the square root of how 
many columns of $\Sigma_{acc}$ are needed to achieve criticality.
The criticality parameter $\mu(t)$ can be derived directly from eq.~\ref{e:sigmat}, as can the ambipolar diffusion timescale $\tau_{AD}$, 
using the fact that $\Sigma = 2\rho_c H$. 

\subsection{Comparison of Timescales}\label{ss:sheettime}

\begin{figure}
  \begin{center}
  \includegraphics[width=\columnwidth]{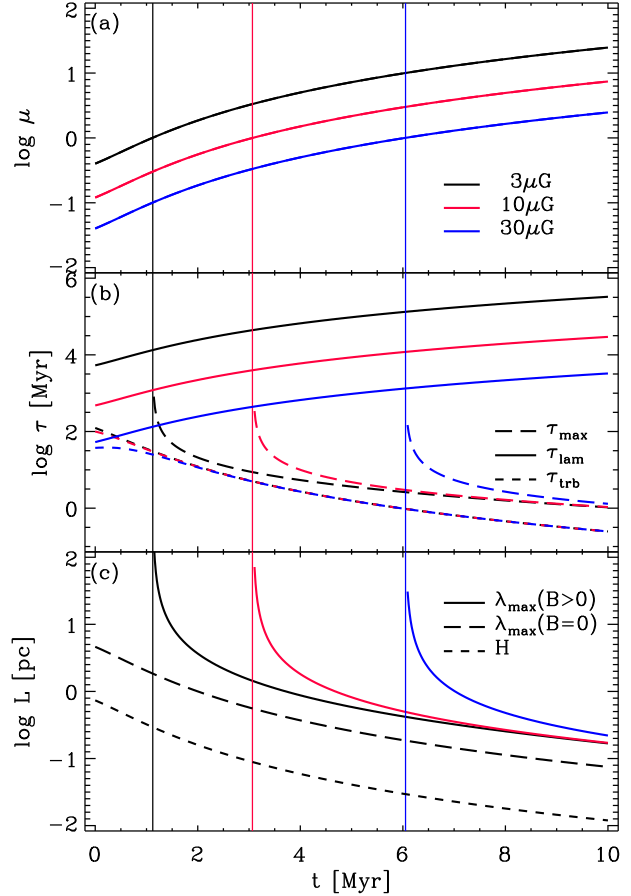}
  \end{center}
  \caption{\label{f:sheet}Time evolution of a sheet accreting at free-fall velocities.
          (a) Magnetic criticality parameter $\mu$ (from eqs.~\ref{e:Ncrit}, \ref{e:sigmat}) for magnetic field 
          strengths $[3,10,30]\mu$G as indicated. Vertical lines denote the time $\tau_{acc}$ (eq.~\ref{e:taccsheet}) 
          at which $\mu=1$.
          (b) Fastest-growing gravitational mode $\tau_{max}$ of a magnetized sheet (long-dashed lines), laminar ambipolar diffusion timescale $\tau_{lam}$ 
          (eq.~\ref{e:tAD}, solid lines), and turbulent diffusion timescale (eq.~\ref{e:tADtrb}, short dashed lines).
          The sheet can become supercritical solely due to accretion on time scales (substantially)
          shorter than the ambipolar diffusion timescales.
          (c) Fastest-growing gravitational length scale $\lambda_{max}$ of a magnetized
          sheet (solid lines, \citealp{1985MNRAS.214..379L}), fastest-growing gravitational mode for an unmagnetized sheet 
          \citep[][long-dashed lines]{1985MNRAS.214..379L}, and
          scaleheight $H$ (eq.~\ref{e:scaleheight}, short dashed lines). The background density is set to $n=100$~cm$^{-3}$, and the 
          reference distance to $z_{ref}=3$~pc.}
\end{figure}

Figure~\ref{f:sheet} summarizes the evolution of an infinite, magnetized sheet accreting diffuse molecular gas 
via gravity
(Sec.~\ref{ss:gravitysheet}). The 
ambient gas density is $n_0=100$~cm$^{-3}$, and the reference distance $z_{ref}=3$~pc. The criticality parameter (panel (a)) is 
proportional to the column density. Vertical lines indicate $\tau_{acc}$ (eq.~\ref{e:taccsheet}). 
Comparing $\tau_{acc}$ to the ambipolar diffusion timescale (panel (b), solid
lines labeled $\tau_{lam}$), we see that $\tau_{acc}\ll \tau_{lam}$. Figure~1 of \citet{2010ApJ...725..466C} suggests that for densities 
$10^3<n<10^4$~cm$^{-3}$, the upper limit for the field strength is between $30$ and $100\mu$G. Yet, such densities would not be expected at early
evolutionary stages (which are of concern here), thus, we do not consider magnetizations higher than $30\mu$G. It is easy to see that cranking
up the magnetic field strength would eventually lead to a regime where $\tau_{lam}<\tau_{acc}$, but such a regime is not physically accessible.
Finally, it is worth pointing out that for $t=0$, there would
be no sheet. The cloud would appear in CO when an $A_V\approx 1$ is reached. For typical flow parameters, this occurs a few Myr after
flow collision \citep{2008ApJ...689..290H}.

The physically correct length scale for laminar ambipolar diffusion would 
be the gravitational forcing scale, i.e. the fastest-growing, gravitationally unstable mode $\lambda_{max}$ of a magnetized sheet 
\citep[][see panel(c), long-dashed lines]{1985MNRAS.214..379L}.
Note that while we show $\lambda_{max}$, the growth rate drops only slowly for wavelengths $\lambda>\lambda_{max}$. For instance,
at $\lambda=5\lambda_{max}$, the timescale is only $\sim 60$\% longer than at $\lambda_{max}$. Thus, we slightly underestimate
the relevance of fragmentation in Figure~\ref{f:sheet}.
Since this length scale is undefined for $\mu\leq 1$, we also show the corresponding hydrodynamical scale 
\citep[][long-dashed lines in panel (c)]{1985MNRAS.214..379L}. 
The short-dashed line shows the scaleheight $H$ (eq.~\ref{e:scaleheight}) for comparison. 
The latter would be an inappropriate choice for an ambipolar diffusion scale in this context, since we are not interested in the {\em vertical}
diffusion of the magnetic field.
Summarizing, {\em the sheet turns supercritical due to accretion 
on timescales substantially shorter than the laminar ambipolar diffusion timescale}. 

Can turbulence accelerate ambipolar diffusion sufficiently to win over accretion? While there is a variety of implementations of turbulent ambipolar diffusion, 
such as acceleration through stagnation point flows \citep{2002ApJ...567..962Z}, stochastic acceleration through field and density
variations \citep{2002ApJ...570..210F}, turbulent mixing generating small scales \citep{2004ApJ...603..165H} and -- to a large extent
combining much of the above -- turbulent compression \citep{2004ApJ...609L..83L}, they all sugest that turbulent ambipolar diffusion can break the
flux-freezing assumption on dynamical rather than on diffusive timescales. While we will focus our discussion on turbulent {\em ambipolar} diffusion
for the sake of consistency with previous literature \citep{2004ApJ...609L..83L,2011ApJ...728..123K}, 
the exact mechanism of how flux-freezing is broken at the smallest scales may not be that relevant, as long as the turbulent diffusivity 
$\lambda_{trb}\gg\lambda_{lam}$, the laminar diffusivity 
\citep[e.g.][see, however, \citealp{2010ApJ...714..442S} for a different point of view]{2002ApJ...578L.113K}. Thus,
$\lambda_{lam}$ could also be due to Ohmic dissipation. Numerical evidence for reconnection diffusion \citep{1999ApJ...517..700L} is given by 
\citet{2013Natur.497..466E}. Thus we will speak of {\em turbulent diffusion} in the following, with the understanding that $\lambda_{lam}=\lambda_{AD}$ 
(see eq.~\ref{e:lamADuse}) in our calculations. 

The short-dashed curves in panel (b) of Figure~\ref{f:sheet} trace the 
evolution of
\begin{equation}
  \tau_{trb} = \frac{H^2}{\lambda_{AD}+H\sigma}.
  \label{e:tADtrb}
\end{equation}
Here we argue that the ambipolar "diffusivity" $\lambda_{AD}$ (eq.~\ref{e:lamADuse}) is enhanced by a turbulent diffusivity of
$\lambda_{trb}=\sigma H$, where $H$ is chosen as 
an outer turbulent scale assuming isotropic turbulence, and $\sigma$ is the turbulent
{\em rms} velocity. 
For the latter, we assume that turbulence within the sheet can be driven by accretion 
\citep{2010A&A...520A..17K}, and can be written as
\begin{eqnarray}
  \sigma &=& \left(4\epsilon\, H\, v_z(H)^2 \dot{\Sigma}/\Sigma\right)^{1/3}\label{e:sigturb}\\
         &=& (4\epsilon)^{1/3} v_z(H)\nonumber\\
         &=& 0.74 \left(\frac{\epsilon}{0.1}\right)^{1/3}\left(\frac{N}{10^{21}\mbox{ cm}^{-2}}\right)^{1/2}\left(\frac{z_{ref}}{\mbox{pc}}\right)^{1/2}\mbox{ km s}^{-1},
         \nonumber 
\end{eqnarray} 
with a driving efficiency $\epsilon=0.1$
\citep[see][for a discussion based on \citealp{2010A&A...520A..17K}]{2013ApJ...769..115H}.
With equations~\ref{e:sigturb} and \ref{e:scaleheight}, the turbulent diffusivity reads
\begin{equation}
  \lambda_{trb} = 3.6\times 10^{22}\left(\frac{\epsilon}{0.1}\right)^{1/3}\left(\frac{N}{10^{21}\mbox{ cm}^{-2}}\right)^{-1/2}\left(\frac{z_{ref}}{\mbox{pc}}\right)^{1/2}
                  \mbox{ cm$^2$ s$^{-1}$}.
  \label{e:lambdatrb}
\end{equation}
The scaleheight $H$ is on the order of a tenth to a few tenths of a parsec.
Comparing the turbulent diffusivity $\lambda_{trb}$ to the laminar ambipolar diffusivity $\lambda_{AD}$ (eq.~\ref{e:lamADuse}), we see that for conditions as to be found in 
molecular clouds, $\lambda_{trb}\gg\lambda_{AD}$. 

From Figure~\ref{f:sheet} we see that although orders of magnitude 
shorter than the laminar ambipolar diffusion timescale $\tau_{lam}$, $\tau_{trb} < \tau_{acc}$ only for the strongest magnetization, which,
however is irrelevant at early times. Also, 
though $\tau_{trb}>\tau_{max}$ for $\sim0.5$~Myr directly after the sheet turns supercritical,
this timespan is much shorter than $\tau_{trb}$ at that point\footnote{The referee points out that the horizontal diffusion timescale may not be the most
important one, for two reasons. First, compression without a horizontal component would leave the field unaffected, and thus horizontal diffusion would be
irrelevant. Yet, this applies only to infall of gas of uniform density. Overdensities could lead to local enhancements, and thus horizontal perturbations. Second,
if there is horizontal compression, the bent field lines would attempt to straighten, leading to horizontal diffusion. This is briefly discussed in App~\ref{a:horizontal}.}.
{\em Thus, we conclude that turbulent diffusion does not dominate the criticality of the sheet.}

\section{Accretion and Ambipolar/Turbulent Diffusion Timescales for a Filament}\label{s:filament}

While for the accreting sheet the choice of field orientation is fairly obvious, and results in $B\propto \rho^0$, 
the situation for an accreting filament is less so. Motivated by observational constraints (see Sec.~\ref{s:introduction}),
we assume a planar field geometry, with the field perpendicular to the filament. Thus, material is free to flow (mostly)
within the field plane onto the filament. Such a geometry would result in $B\propto \rho^0$, and an accretion rate of $\sim 4\rho_0 v_0 H$, where
$H$ is the scaleheight of the filament. It may be less restrictive to assume that
the inflow is not restricted to the plane, but extends over the whole azimuthal range. If mass and flux are conserved during
radial contraction, then $B\propto \rho^{1/2}$, a scaling that is also more consistent with the observed 
increase in field strength at higher densities 
\citep{2010ApJ...725..466C}.

\subsection{Model}
We follow the evolution of a filament under free-fall, steady-state accretion, using a
model series of externally pressurized, accreting cylinders, with $B\propto \rho^{1/2}$ \citep{2013ApJ...776...62H}. Assumptions, derivations and
expressions can be found in that paper. Since we envisage the filaments to be embedded in a flattened, evolved molecular cloud, we set the ambient
density to $100$~cm$^{-3}$, corresponding to diffuse molecular gas. Infall velocities range around a km~s$^{-1}$ \citep[see Fig.~1 of ][]{2013ApJ...776...62H}.
We calculate two criticality parameters: The radial criticality parameter 
\begin{equation}
  \mu_{rad}\equiv \frac{m}{m_{rad}}
  \label{e:mcritrad}
\end{equation}
compares the filament's line mass $m$ with the critical magnetic line mass, 
\begin{equation}
  m_{rad} = 0.24\frac{BR}{\sqrt{G}}+1.66\frac{c_s^2}{G},
  \label{e:mcmag}
\end{equation}
as determined numerically by \citet{2013prpl.conf1S026T}.
For the longitudinal criticality parameter, we adapt the expressions given by \citet{1983A&A...119..109B} for the number of Jeans masses in 
a magnetized, prolate ellipsoid with axes $a>b=c$,
\begin{equation}
  \mu_{lon}\equiv \frac{\pi G \rho c^2}{15 e c_s^2}\ln\left(\frac{1+e}{1-e}\right),
  \label{e:mcritlon}
\end{equation}
where the ellipticity $e$ is defined by $c^2=a^2(1-e^2)$. We use an aspect ratio of $a/c=10$. Larger aspect ratios
increase $\mu_{lon}$, thus, we can consider our choice as a lower bound.
As \citet{1983A&A...119..109B} discusses, the limit towards high ellipticity (infinitely long filaments) yields
an infinite Jeans number and thus does not converge to the radial expression, since these are different collapse modes.
Ambipolar diffusion timescales are calculated as above. The wavelength of the fastest-growing mode
depends on whether the filament resides in a vacuum \citep[infinite radial extent, or an infinite overpressure, ][]{1964ApJ...140.1056O}, or whether
a ``truncated'' \citep[or pressurized,][]{2012A&A...547A..86F} filament is considered. We show both cases for comparison, using 
the expressions given by \citet{1987PThPh..77..635N} for the infinite cylinder, and the polynomial fits by \citep{2012A&A...547A..86F} for the 
pressurized cylinder. For the sake of simplicity, we forego the discussion of varying external pressure due to accretion \citep{2013ApJ...776...62H}. 

\subsection{Comparison of Timescales}
\begin{figure}
  \begin{center}
  \includegraphics[width=\columnwidth]{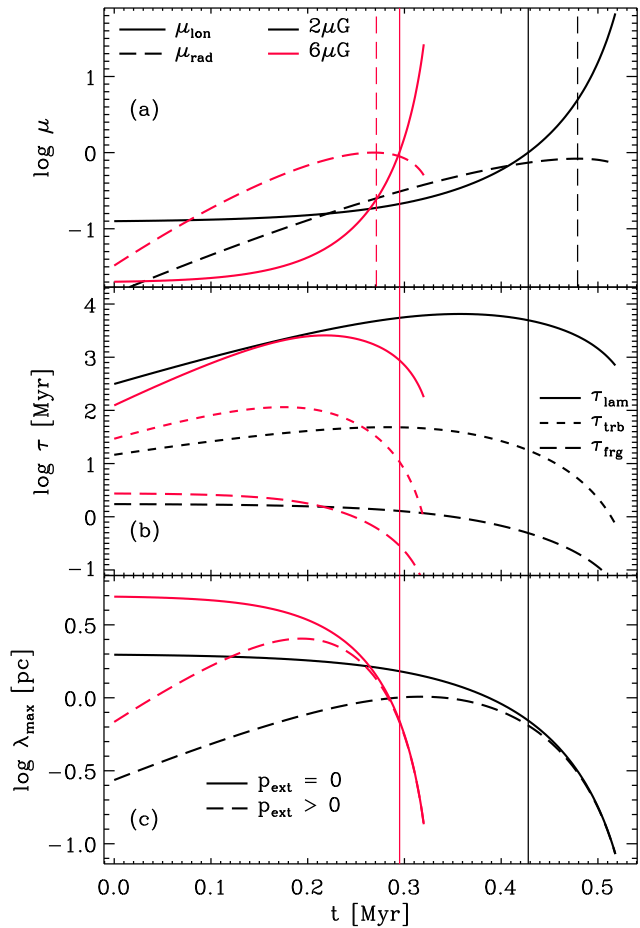}
  \end{center}
  \caption{\label{f:filament}
           Time evolution of a filament accreting at free-fall velocities.
           (a) Longitudinal (solid lines, eq.~\ref{e:mcritlon}) and radial (dashed lines, eq.~\ref{e:mcritrad}) 
           criticality parameters. 
           Longitudinal and radial criticality (indicated by vertical lines) are reached at about the same time, with longitudinal criticality winning
           over radial for weak magnetizations. Thus, material will collect first {\em along} the filament.
           (b) Laminar ambipolar diffusion timescale (solid lines), turbulent diffusion timescale (dashed lines), for length scales set by the
           gravitationally most unstable mode (panel (c)). 
           Fragmentation timescales (long-dashed lines) are substantially shorter than ambipolar diffusion timescales at time criticality is reached.
           (c) Length scales of gravitationally most unstable mode. We distinguish between infinite (solid lines) and finite (dashed lines)
           cylinders. They converge for large line masses, or large overpressures \citep{2012A&A...547A..86F}.}
\end{figure}

Figure~\ref{f:filament} summarizes the evolution of a filament accreting at 
free-fall velocities, assuming $B\propto \rho^{1/2}$. 
Results are shown for two magnetizations, corresponding to {\em initial} field strengths
of $2$ and $6\mu$G. During the evolution of the filament, fields $\gtrsim 100\mu$G are reached (see also Sec.~\ref{s:discussion}).

Evolution timecales are below one Myr overall. The filaments reach criticality after a fraction of a Myr (panel (a)), with global 
longitudinal criticality $\mu_{lon}$ (eq.~\ref{e:mcritlon}) close to radial criticality $\mu_{rad}$ (eq.~\ref{e:mcritrad}), and 
winning for weak magnetizations. 
Panel (b) shows that the fragmentation timescales drop to values close to the longitudinal accretion timescale. 
The laminar ($\tau_{lam}$, solid lines) and turbulent diffusion timescales ($\tau_{trb}$, dashed lines) are all longer than the time to 
reach (longitudinal) criticality, and they are all
larger than $10$~Myr at $\tau_{acc}$, although they do drop for times $>\tau_{acc}$. Yet that drop is irrelevant for our
purposes: the filament already has grown critical due to accretion. {\em Thus, as in the sheet case, filaments
are likely to reach criticality through accretion, before ambipolar diffusion can have a significant effect.} 
Note that the timescales in panel (b) refer to externally pressurized (truncated) cylinders
\citep{2012A&A...542A..77F}. The corresponding timescales for cylinders in vacuum \citep{1964ApJ...140.1056O,1987PThPh..77..635N} 
differ only for small evolution times, being always
larger than for truncated cylinders (only shown in Fig.~\ref{f:filament} for the length scales).
Panel (c) shows the length scales of the fastest growing mode, derived for infinite \citep[][solid lines]{1987PThPh..77..635N} and
pressurized \citep[][dashed lines]{2012A&A...547A..86F}. As the latter authors discuss, the expressions converge for large line masses, i.e.
larger overpressures.

\section{Discussion}\label{s:discussion}

We compared ambipolar diffusion and accretion timescales for sheets (Sec.~\ref{s:sheet}) and filaments (Sec.~\ref{s:filament}), finding
that in both geometries, accretion will dominate the evolution from sub- to supercritical structures. For the sheets, we considered
two cases; the assembly of a molecular cloud from the diffuse (atomic) interstellar medium, and the evolution of a dense cloud due to
accretion of ambient molecular gas. For the filament geometry, we only consider gravitationally driven accretion. 

Our assumption of free-fall 
accretion may seem overly extreme.  
Yet, already order-of magnitude estimates (eqs.~\ref{e:tauc}, \ref{e:tAD}) with constant (rather low) velocities suggest 
that accretion dominates over ambipolar diffusion. Also, accretion velocities are at most on the order of a km~s$^{-1}$ 
\citep{2013ApJ...769..115H,2013ApJ...776...62H}, consistent with observed 
values \citep{2013ApJ...766..115K,2013MNRAS.436.1513F}.

\subsection{Sheets}

\subsubsection{Choice of Length Scales} 
The choice of length scales is somewhat open, yet not completely arbitrary. We are interested in motions perpendicular to the magnetic
field. The only characteristic length available in the sheet plane is that of the gravitationally
most unstable mode (this also would be the length scale at which the strongest forces occur). Shorter length scales will not be unstable yet,
and longer scales would grow more slowly. One could argue that substructure in the sheet could reduce the length scale, and thus the
laminar ambipolar diffusion timescale. Yet (1) the substructure would have to be non-linear, specifically in a finite sheet, to win over large-scale collapse
modes \citep{2004ApJ...616..288B,2012ApJ...756..145P}, and (2) this case is addressed by the turbulent diffusion timescale. Note that for the latter,
the ``forcing scale'' is still set by the gravitationally most unstable mode. 
Such an interpretation is also consistent with the findings
of \citet{2004ApJ...603..165H}, that turbulent ambipolar diffusion is not a flux transport mechanism, but acts through breaking the flux-freezing assumption
locally. 

\begin{figure*}
  \begin{center}
  \includegraphics[width=\textwidth]{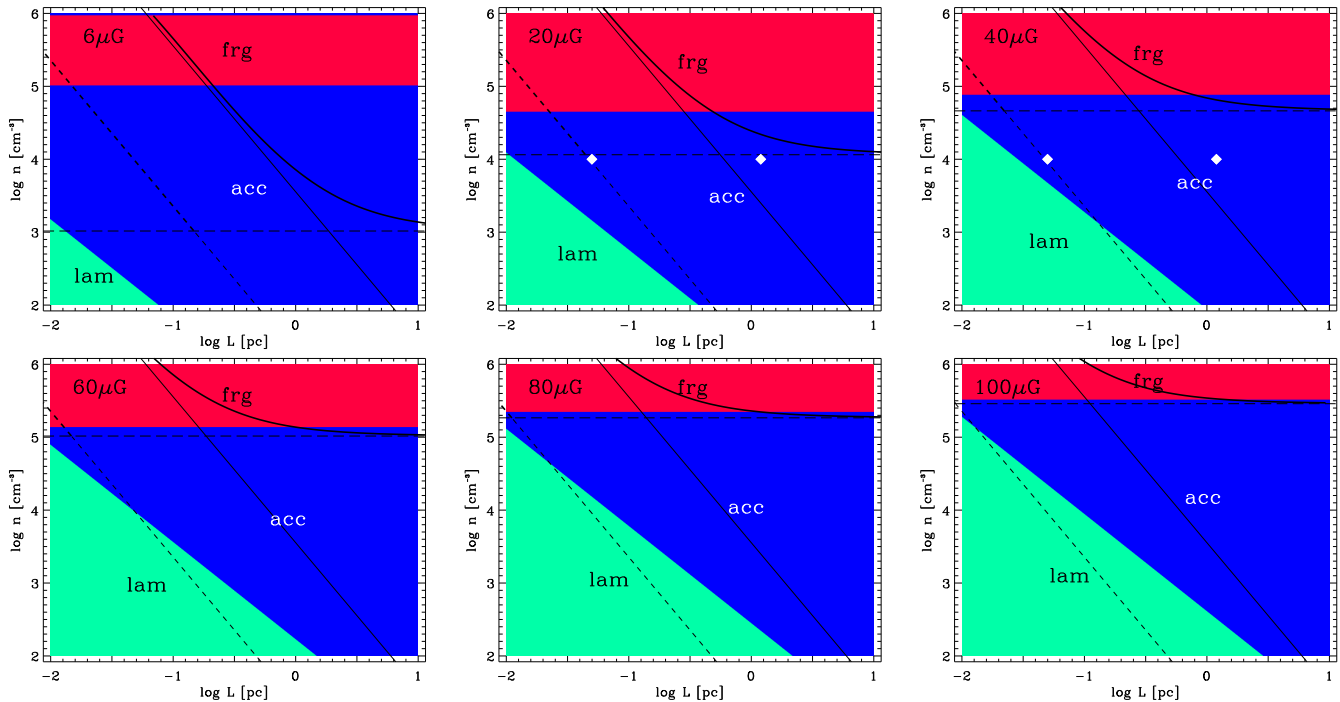}
  \end{center}
  \caption{\label{f:sheetmaptrb0}
          Map of the shortest timescales in the length scale-density plane for a sheet. For a given combination of $(L,n)$,
          all relevant timescales are compared, and the regime of the shortest timescale is color-coded as blue for
          accretion ({\em acc}), red for gravitational fragmentation ({\em frg}), and light green for laminar ambipolar diffusion ({\em lam}).
          The solid curved line
          shows the wavelength of the most unstable (magnetized) mode, $\lambda_{max}$, against the sheet's central density \citep{1985MNRAS.214..379L},
          the solid straight line is the corresponding hydrodynamical mode, and the dashed
          line shows the sheet's scaleheight against central density. Criticality is indicated by the
          horizontal, long-dashed line (eq.~\ref{e:rhocrit}). 
          White diamonds show the initial conditions for models run by Kudoh \& Basu (\citeyear{2011ApJ...728..123K}, see Sec.\ref{sss:kudohbasu}).}
\end{figure*}

\begin{figure*}
  \begin{center}
  \includegraphics[width=\textwidth]{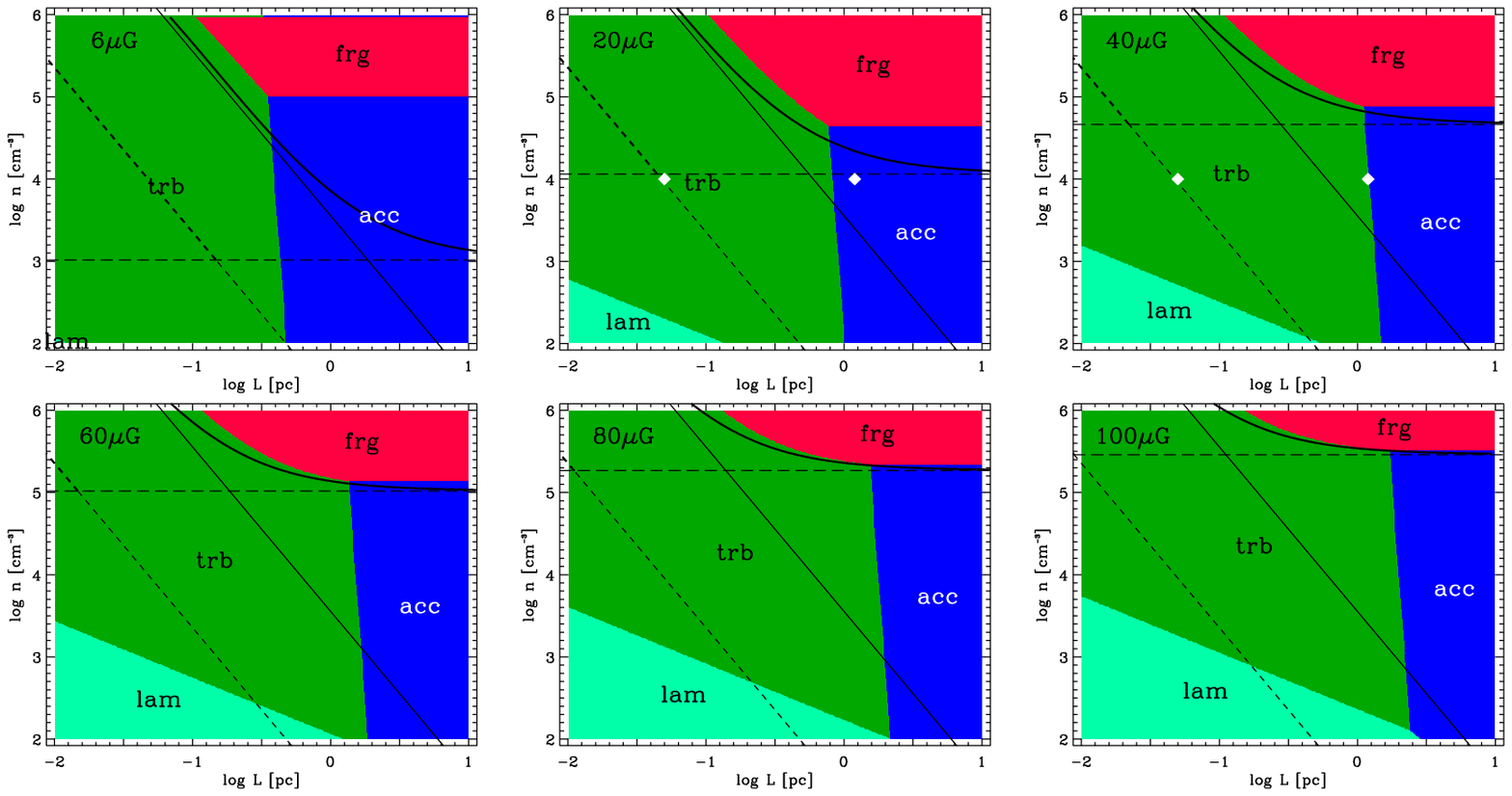}
  \end{center}
  \caption{\label{f:sheetmaptrb1}Same as Figure~\ref{f:sheetmaptrb0} including the turbulent diffusion timescale.
          Map of the shortest timescales in lengthscale-density plane, for a sheet. For a given combination of $(L,n)$,
          all relevant timescales are compared, and the regime of the shortest timescale is color-coded as blue for
          accretion ({\em acc}), red for gravitational fragmentation ({\em frg}), light green for laminar ambipolar diffusion ({\em lam}), 
          and dark green for turbulent diffusion ({\em trb}). The solid curved line
          shows the wavelength of the most unstable (magnetized) mode, $\lambda_{max}$, against the sheet's central density \citep{1985MNRAS.214..379L},
          the solid straight line is the corresponding hydrodynamical mode, and the dashed
          line shows the sheet's scaleheight against central density. Criticality is indicated by the
          horizontal, long-dashed line (eq.~\ref{e:rhocrit}). 
          White diamonds show the initial conditions for models run by Kudoh \& Basu (\citeyear{2011ApJ...728..123K}, see Sec.\ref{sss:kudohbasu}).}
\end{figure*}

Since the choice of length scales may be contentious (and is relevant for the interpretation of our results), we highlight its role 
in Figures~\ref{f:sheetmaptrb0} and \ref{f:sheetmaptrb1}. 
They map out the regimes of the shortest (dominant) timescale in the length-density plane ($\log L,\log n$). 
Colors stand for timescales, indicated by {\em acc} for accretion (blue), {\em frg} for fragmentation (red), {\em lam} 
for laminar ambipolar diffusion (light green), and (for Fig.~\ref{f:sheetmaptrb1}) 
{\em trb} for turbulent diffusion (dark green). The fragmentation timescale
\citep[][see Sec.~\ref{ss:sheettime}]{1985MNRAS.214..379L} varies following the full dispersion relation.
The turbulent diffusion timescale is calculated using the length
scale on the $x$-axis. We chose to present two versions of these maps, one (Fig.~\ref{f:sheetmaptrb0}) without, and the other with turbulent  
diffusion (Fig.~\ref{f:sheetmaptrb1}). Each Figure contains maps for six magnetic field strengths 
($6,20,40,60,80,100\mu$G). In each map, the solid line traces the gravitationally most unstable mode derived from 
\citet[][see Appendix~\ref{a:sheet}]{1985MNRAS.214..379L},
and the short dashed line stands for the scaleheight of the sheet (eq.~\ref{e:scaleheight}). 
Criticality is indicated by the horizontal, long-dashed line, combining equations~\ref{e:Ncrit} and \ref{e:scaleheight}, and the expressions for the central
density in an isothermal sheet, $\rho_0 = \Sigma/(2H)$, to 
\begin{equation}
  \rho_{crit} = \frac{B^2}{8\pi c_s^2}.
  \label{e:rhocrit}
\end{equation}
In other words, the critical central density is given by comparing the magnetic and thermal pressure, under the assumption
that the field is perpendicular to the disk. 

Figures~\ref{f:sheetmaptrb0} and \ref{f:sheetmaptrb1} can be used in various ways. First, we can ask for which combinations of length scale and
central density ambipolar diffusion might be relevant. This alleviates the problem of having to fix the length scale in our earlier discussion.
Second, one can easily read off an evolutionary sequence (via the density-scaleheight relation, eq.~\ref{e:scaleheight}) 
from a subcritical to a critical state for a given field strength,
and check whether the sheet gets supercritical due to accretion before reaching the ambipolar diffusion regime (laminar or turbulent).
It is worth pointing out that the indicated field strength refers to the diffuse ISM, {\em if} the 
lines for $H(n)$ and $\lambda_{max}(n)$ are interpreted as evolutionary sequences, since under our assumption $B\propto\rho^0$. 
The density of the inflows is kept constant at $n=100$~cm$^{-3}$. 

\subsubsection{Model Comparison}\label{sss:kudohbasu}
In a series of papers, \citet{2008ApJ...679L..97K,2009RMxAC..36..278K,2011ApJ...728..123K} 
explored the gravitational fragmentation of infinite, magnetized, 
sheet-like clouds mediated by ambipolar diffusion. In their most recent work, they find that laminar ambipolar diffusion results in a fragmentation timescale
of a few $10^7$ years, while the presence of moderately supersonic flows can reduce the fragmentation timescale by a factor
of $~10$. \citet{2007RMxAA..43..123B} discuss that the former timescale is inconsistent with local observations
of stellar age spreads of young stars in molecular clouds. Yet, leaving that issue aside for the moment, we attempt to put 
the results of Kudoh \& Basu in the context of our timescale discussion. 

\citet{2011ApJ...728..123K} quote characteristic values (assuming a sound speed and a density) for their models. Specifically, at a sound
speed of $0.2$~km~s$^{-1}$, identical to what we use in our discussion, and a central sheet density of $n=10^4$~cm$^{-3}$, their choice of $\beta=1.0$
results in a field strength of $B=20\mu$G, and that of $\beta=0.25$ in  $B=40\mu$G (this would correspond to their model V2). 
We indicate two length scales in the corresponding
panels ($20,40\mu$G) of Figures~\ref{f:sheetmaptrb0} and \ref{f:sheetmaptrb1} by white diamonds: The smaller length scale is the scaleheight
of the sheet, and
the larger length scale is the scale of the largest turbulent mode in their simulation. 

A glance at Figure~\ref{f:sheetmaptrb0} (for $B=20,40\mu$G) highlights the situation for laminar ambipolar diffusion: even at a scale corresponding to the 
scaleheight of the sheet ($0.05$~pc for the parameters above), $\tau_{acc}<\tau_{AD}$, and thus, the sheet will become critical due to 
accretion before ambipolar diffusion can have a significant effect. Following the dashed line (i.e. the scale height in dependence of density) towards 
lower densities can be used as a (reverse) evolutionary sequence for a mass-accreting sheet (and thus increasing $\Sigma$ and $\rho$).
For $B=40\mu$G, only sheets with central densities $\lesssim 10^3$~cm$^{-3}$ are dominated
by ambipolar diffusion on scales of $H$. For larger scales, accretion still wins. For $B=20\mu$G, there is no viable range of densities for ambipolar 
diffusion to be dominant.

Turbulent diffusion (Fig.~\ref{f:sheetmaptrb1}) will dominate over accretion on length scales of the scale height $H$ at $20$ and $40\mu$G (and 
for any higher magnetization), yet, the scale of the fastest growing mode as well as the outer turbulent scale of the simulations by 
\citet{2011ApJ...728..123K} are still residing in the accretion-dominated (blue) regime. It should be noted that the "turbulent diffusion" discussed
here assumes a different mechanism than the ambipolar diffusion acceleration in Kudoh \& Basu's work. Here, we assume that the laminar ambipolar diffusion 
rate can be 
accelerated by a turbulent diffusivity \citep[see, e.g., ][]{2002ApJ...578L.113K}. 
\citet{2011ApJ...728..123K} rely on (shock) compressions due to supersonic turbulence,
leading to steep gradients and thus to a local acceleration of ambipolar diffusion in the shocked regions \citep[see also][]{2004ApJ...609L..83L}.
Whether turbulent diffusion really wins over accretion to make the sheet supercritical probably is best tested with a simulation along the lines
of \citet{2011ApJ...728..123K}, but adding mass to the (sheet-like) cloud.

\subsection{Filaments}

\subsubsection{Choice of Length Scales}
\begin{figure*}
  \begin{center}
  \includegraphics[width=\textwidth]{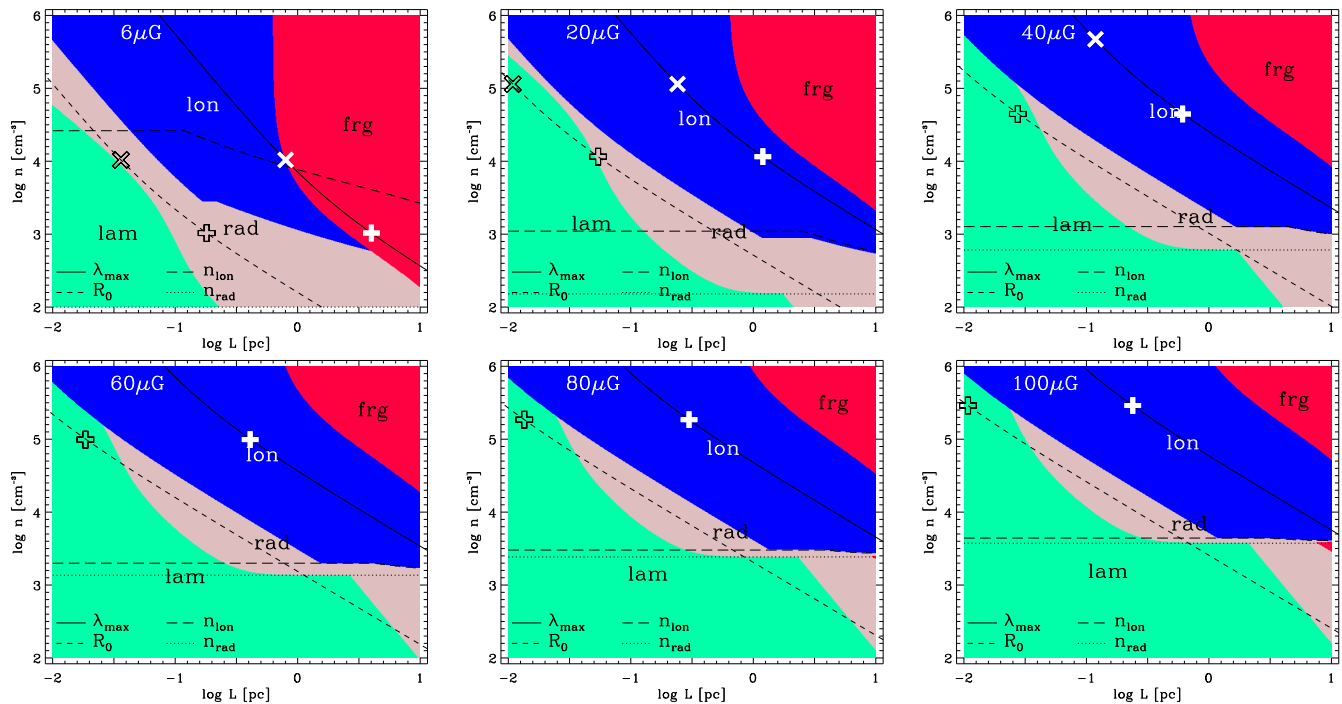}
  \end{center}
  \caption{\label{f:filmaptrb0}Map of the shortest timescales in the lengthscale-density plane, for a filament. For a given combination of $(L,n)$, 
          all relevant timescales are compared, and the regime is color-coded as longitudinal criticality (blue, {\em lon}), radial criticality 
          (pink, {\em rad}), gravitational fragmentation (frg, {\em red}), and laminar ambipolar diffusion (light green, {\em lam}). 
          The solid line shows the wavelength of the most unstable mode, $\lambda_{max}$ \citep{1987PThPh..77..635N} 
          against the filament density, and the dashed
          line shows the filament's core-radius $R_0$ (eq.~\ref{e:R0}). Longitudinal criticality is indicated by the 
          long-dashed line (eq.~\ref{e:mcritlon}), and radial criticality by the dotted line. 
          Symbols denote the evolution of the two filament accretion models shown in Fig.~\ref{f:filament}.}
\end{figure*}

\begin{figure*}
  \begin{center}
  \includegraphics[width=\textwidth]{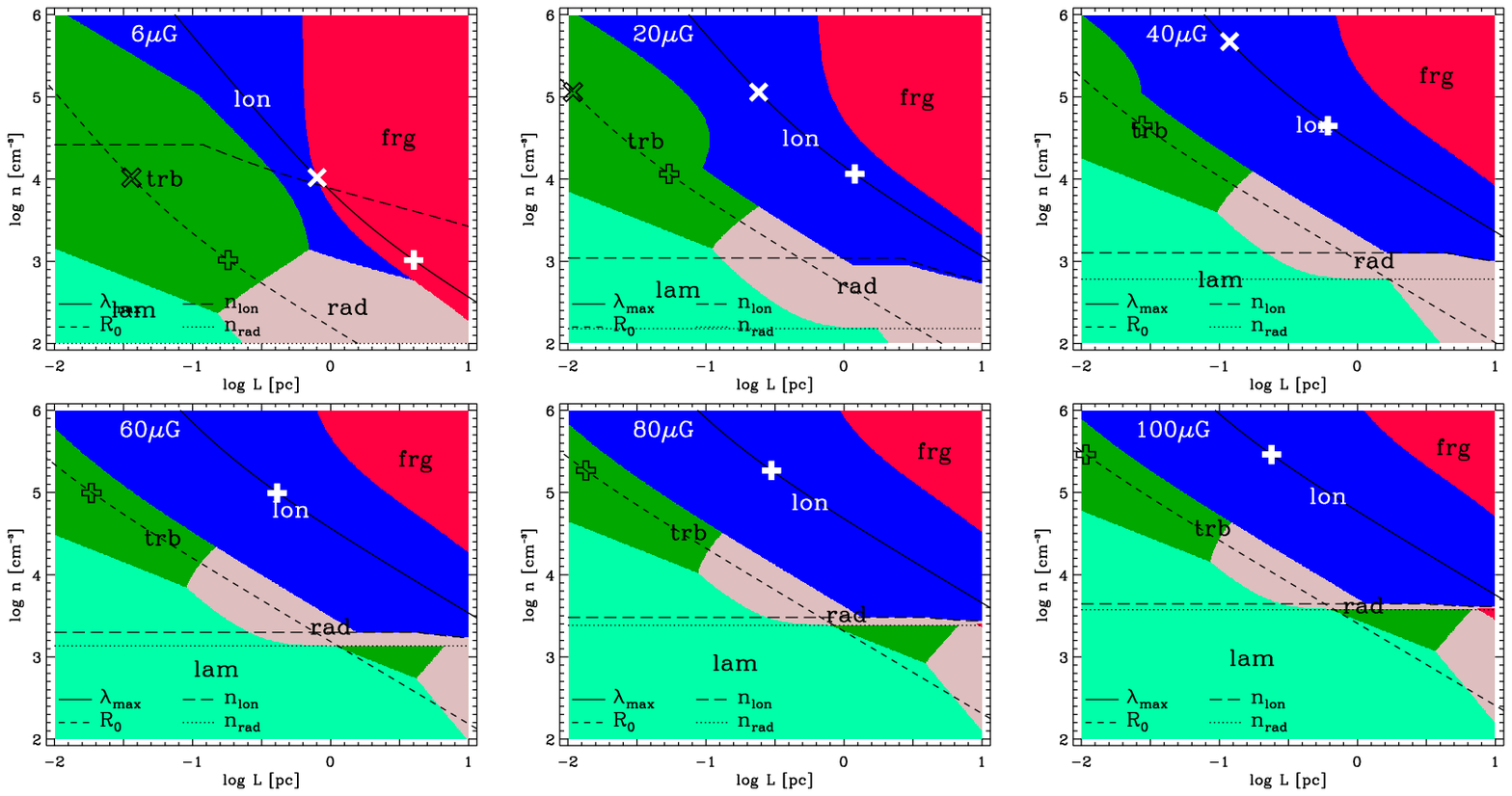}
  \end{center}
  \caption{\label{f:filmaptrb1}Same as Fig.~\ref{f:filmaptrb0}, including turbulent diffusion.
          Map of the shortest timescales in the lengthscale-density plane, for a filament. For a given combination of $(L,n)$, 
          all relevant timescales are compared, and the regime is color-coded as longitudinal criticality (blue, {\em lon}), radial criticality 
          (pink, {\em rad}), gravitational fragmentation (frg, {\em red}), laminar 
          ambipolar diffusion (light green, {\em lam}), and turbulent  (dark green, {\em trb}).
          The solid line shows the wavelength of the most unstable mode, $\lambda_{max}$ \citep{1987PThPh..77..635N} 
          against the filament density, and the dashed
          line shows the filament's core-radius $R_0$ (eq.~\ref{e:R0}). Longitudinal criticality is indicated by the 
          long-dashed line (eq.~\ref{e:mcritlon}), and radial criticality by the dotted line. 
          Symbols denote the evolution of the two filament accretion models shown in Fig.~\ref{f:filament}.}
\end{figure*}

Figures~\ref{f:filmaptrb0} and \ref{f:filmaptrb1} highlight the effect of setting a specific length
scaley by mapping the shortest timescales in the $(L,n)$-plane. There are two
characteristic scales now (for an infinite filament): the radial extent, here described by the scale radius 
\begin{equation}
  R_0^2 = \frac{2c_s^2}{\pi G\rho_c}\label{e:R0}
\end{equation}
\citep{1964ApJ...140.1056O}, and the length scale of the 
gravitationally most unstable mode along the filament. Both are marked in the Figures by short-dashed and solid lines, respectively. The 
longitudinal timescale ({\em lon}, blue) refers to the global criticality (eq.~\ref{e:mcritlon}) of a {\em finite} filament at a given length 
indicated by $L$ (x-axis). 
Following Bastien's (\citeyear{1983A&A...119..109B}) analysis, we assume that the filament must have an aspect ratio of $>2$ for longitudinal 
collapse to be viable. Chosing a density $n$ (y-axis) results in a filament scale radius $R_0$ (dashed line). Together with an assumed 
filament length, results for any aspect ratio can be 
read off. The longitudinal fragmentation timescale ({\em frg}, red) refers to the gravitationally most unstable mode \citep{1987PThPh..77..635N} for an infinite filament.

The time evolution of a filament maps slightly differently to these plots, compared to the sheet case. There, the magnetic field strength does not change, whereas, here, 
$B\propto \rho^{1/2}$. To help understanding the time sequence, we overplotted the positions of the two models shown in Figure~\ref{f:filament} 
in $(L,n)$-space, for each magnetic fieldstrength.  We use two characteristic scales, the most
unstable mode (filled symbols), and the scale radius $R_0$ (eq.~\ref{e:R0}, open symbols), to highlight the role of the scale choice.
We note two issues: (a) Evolving from weak to strong fields, all filaments (in the laminar case) on scales between $0.1$ and $10$~pc 
start out in a region where criticality is reached faster by accretion than by ambipolar diffusion, if we interpret $R_0$ and $\lambda_{max}$ as a lower and upper
bound for reasonable physical scales. Only at higher magnetic fields (later in the evolution), some of the models enter the ambipolar diffusion regime ({\em lam},
light green) on scales of $R_0$. Yet, at those densities, the filament is already radially and longitudinally supercritical (the densities reside
above the long-dashed and dotted lines).
(b) With increasing field strength, the lines for longitudinal and radial criticality (long dashed and dotted) move
to higher densities, yet, if we interpret the sequence of magnetic fields as evolutionary sequence, we conclude that the filaments evolve faster
to higher densities than the critical density with increasing field strength. {\em Taken together, these two observations 
show that even if the filament
started out subcritical, it is driven to supercriticality by accretion rather than by laminar ambipolar diffusion.}

A fair fraction of the small scales is residing in the
regime dominated by turbulent diffusion (Fig.~\ref{f:filmaptrb1}, {\em trb}, dark green). Similarly as in the discussion of turbulent diffusion 
in sheets, it remains to be seen what the 
actual characteristic length scales are. Turbulent diffusion does not play a role on the scale of the most unstable mode, but it may play a role
on scales a few times the core radius (few tenths of a parsec). Overall, with increasing field strength, the regime of turbulent diffusion dominance
recedes to higher densities and smaller scales. 

Another way to see this is that longitudinal collapse (fragmentation) dominates until the aspect ratio approaches $2$, i.e. once the filament has 
evolved (fragmented) into a pre-stellar core. While formally we enter the turbulent diffusion regime at that point, it should be noted that for all
practical purposes, the densities are well above the longitudinal and radial critical density, i.e. the structure is already supercritical and
does not need any help from ambipolar diffusion. 

\section{Summary}
Our discussion of the dominant timescales in accreting, magnetized sheets and filaments is motivated by two observations. First, 
magnetic fields seem to be fairly well-ordered around filamentary molecular clouds \citep{2008A&A...486L..13A,2013A&A...550A..38P,2013MNRAS.436.3707L}, 
suggesting large-scale (accretion) flows along the field lines. Evidence for gas accretion onto filaments is mounting 
\citep{2013ApJ...766..115K,2013MNRAS.436.1513F}. Second, while the diffuse interstellar medium is generally
magnetically subcritical, the dense gas ($n\gtrsim 300$~cm$^{-3}$) seems to be on average slightly supercritical by a factor of $2$ 
\citep{2010ApJ...725..466C}. Two ideas have been suggested for the transition between sub- and super-criticality: ambipolar diffusion 
(based on \citealp{1956MNRAS.116..503M}, more recently \citealp{2008ApJ...679L..97K,2011ApJ...728..123K}),
and accretion \citep{1995ApJ...441..702V,2001ApJ...562..852H}. We explore the role of ambipolar diffusion and accretion to drive 
an initially subcritical sheet or filament 
to criticality, by comparing the relevant timescales.\\

\noindent{\bf(1)} While ambipolar diffusion is present, it does not affect the evolution of an accreting sheet or filament. The accretion
timescales are substantially shorter in the subcritical regime, rendering sheets (Fig.~\ref{f:sheetmaptrb0}) and filaments (Fig.~\ref{f:filmaptrb0})
supercritical before laminar ambipolar diffusion can affect the mass-to-flux ratio substantially. This agrees with the findings of 
\citet{2011ApJ...728..123K}, who quote timescales of a few $10^7$ years to drive a sheet to criticality via laminar ambipolar diffusion. 
For sheets, increasing densities reduce the importance of laminar ambipolar diffusion, while for filaments, 
the ambipolar diffusion regime can be re-entered for high densities and small 
scales, yet at a stage when the filament has turned substantially supercritical.\\

\noindent{\bf(2)} Turbulent diffusion may be important in a sheet-like geometry 
(Fig.~\ref{f:sheetmaptrb1}) for strong magnetizations ($B>20\mu$G). Yet, such a field
strength would be a factor $2$ to $4$ higher than observed in the diffuse interstellar gas. In other words, the strong magnetizations refer to a stage of
the cloud when the field has been amplified well beyond its background level -- by gravity. For filaments (Fig.~\ref{f:filmaptrb1}), turbulent  
diffusion plays only a role for small scales -- and only, once the filament has been driven to criticality by accretion.\\

Note that we are {\em not} arguing that all regions of molecular clouds are magnetically supercritical.  Rather,
our analysis shows that the densest regions of such clouds wherein star formation takes place will generally become
supercritical due to the accumulation of mass -- that is, as a result of their formation -- before ambipolar diffusion becomes important.
Magnetic fields may still be important globally in reducing the efficiency of star formation. 
In fact \citet{2010ApJ...725..466C} point out that some clouds may well be magnetically dominated, and in any case the
mean mass-to-flux ratio they infer in dense regions is only a factor of two above criticality, suggesting
that magnetic forces do play a dynamical role \citep[see][for how supercritical magnetic fields affect the gravitational
fragmentation and collapse]{2001ApJ...547..280H,2005ApJ...630L..49V,2008MNRAS.385.1820P}. 
\citet{2012ApJ...755..130M} demonstrate with infrared polarimetry mapping how strongly the mass-to-flux ratio can vary
across a single cloud (see their Fig.~14).

We conclude that the effects of accretion 
must be considered in any analysis of the dynamical state of dense star-forming gas.
We further find that
laminar or turbulent diffusion -- while present -- is unlikely to control the evolution of a cloud
from subcritical to supercritical. 

\section*{Acknowledgements}
We thank the referee for a thought-provoking report. FH acknowledges support by UNC-CH. LH acknowledges partial support by
UM.

\appendix

\section{Ambipolar Diffusion Timescale}\label{a:adtime}
The ambipolar diffusion timescale over a length scale $L$ at an ambipolar diffusivity $\lambda_{AD}$ is given by
\begin{equation}
  \tau_{AD} = \frac{L^2}{\lambda_{AD}}.
\end{equation}
Following \citet{2003ApJ...583..229H}, we write $\lambda_{AD}$ in dependence of the field strength $B$, the ion and
neutral particle densities $n_i$ and $n_n$, and the molecular weights $\mu_i=29$ and $\mu_n=2$, assuming a single species of HCO$^+$ and 
H$_2$ each -- the exact choices do not affect the results. For instance, choosing C$^+$ and H for diffuse clouds results in 
$\mu_i\mu_n/(\mu_i+\mu_n)=0.92$ instead of $1.87$. We also need  the
ionization fraction $x_i\equiv n_i/n_n$, and the rate coefficient for elastic collisions 
$\langle\sigma v\rangle=1.5\times^10{-9}$~cm$^3$~s$^{-1}$ \citep{1983ApJ...264..485D}. Taken together, this results in 
\begin{equation}
  \lambda_{AD} = \frac{\mu_i+\mu_n}{4\pi\langle\sigma v\rangle\mu_i\mu_n m_H x_i} \left(\frac{B}{n_n}\right)^2.
  \label{e:lamAD}
\end{equation}
Assuming ionization-recombination balance, the ionization degree can be approximated by
\begin{equation}
  x_i = 1.2\times 10^{-5} n_n^{-1/2}
  \label{e:iondeg}
\end{equation}
\citep{1979ApJ...232..729E,1980PASJ...32..405U}.
Then the ambipolar diffusivity can be written (for molecular cloud conditions) as
\begin{equation}
  \lambda_{AD} = 8.2\times 10^{16} \left(\frac{B}{5\mu\mbox{G}}\right)^2\left(\frac{n}{300\mbox{ cm}^{-3}}\right)^{-3/2} \mbox{ cm}^2\mbox{s}^{-1},
  \label{e:lamADuse}
\end{equation}
and the ambipolar diffusion timescale reads
\begin{equation}
  \tau_{AD} = 1.38\times 10^3 \left(\frac{L}{\mbox{pc}}\right)^2\left(\frac{n}{3\times 10^2\mbox{cm}^{-3}}\right)^{3/2}\left(\frac{B}{5 \mu\mbox{G}}\right)^{-2}\mbox{ Myr}.
  \label{e:tauADapp}
\end{equation}

We note that equation~\ref{e:iondeg} applies to conditions within dense molecular clouds, with cosmic rays as the only ionization source. Thus, equation~\ref{e:tauADapp}
tends to {\em underestimate} the ionization degree for diffuse gas, specifically for conditions with $A_V<1$. In other words, until the ambient UV field has been 
shielded, ambipolar diffusion will be irrelevant. From equation~\ref{e:Ncrit}, we see that the shielding column density corresponds to a field strength of $\approx 8\mu$G. 

\section{Neglecting the ambient gas column}\label{a:nocolumn}

Equation~\ref{e:vzsheet} is only valid if the column density does not change appreciably while a single column out to $z_{ref}$ is being accreted, or 
if $\Sigma_{acc}\ll\Sigma$. If we assume that $z_{ref}\gg H$, the argument is simplified, while considering the most
unfavorable case. The ambient column then can be written as 
\begin{equation}
  N_{acc}=\frac{\Sigma_{acc}}{\mu m_H}=6.2\times10^{18}\left(\frac{z_{ref}}{\mbox{pc}}\right)\left(\frac{n_0}{\mbox{cm}^{-3}}\right)\mbox{ cm}^{-2}.
\end{equation}
The constant term $b$ (eq.~\ref{e:bpar}) in equation~\ref{e:sigmat} corresponds to the equivalent column of an isothermal sheet of scale height $z_{ref}$, and
can be written as
\begin{equation}
  \frac{b}{\mu m_H} = 1.58\times10^{20}\left(\frac{c_s}{0.2\mbox{ km s}^{-1}}\right)\left(\frac{z_{ref}}{\mbox{pc}}\right)^{-1} \mbox{ cm}^{-2}
\end{equation} 
for $\mu=2.36$.
Thus, $N_{acc}$ is generally negligible. 

\section{Neglecting ram pressure of inflows}\label{a:lowpress}
By setting the lower accretion bound to the scaleheight of the sheet, we assumed that the ram pressure of the infalling gas, 
$p_{ram} = \rho_0 v_z(H)^2$, can be neglected, and thus does not compress the sheet noticeably. We show this by comparing $p_{ram}$ to the
internal sheet pressure, $p_{int} = \pi G \Sigma/2$. Then,
\begin{equation}
  \frac{p_{ram}}{p_{int}}=2.5\times 10^{-2} \left(\frac{n_0}{\mbox{cm}^{-3}}\right)\left(\frac{z_{ref}}{\mbox{pc}}\right)\left(\frac{N}{10^{21}\mbox{cm}^{-2}}\right)^{-1}.
\end{equation}
Here, we have assumed that $H/z_{ref}\ll 1$, an assumption that actually increases $p_{ram}$. Thus, the above expression is a conservative estimate. It also should be 
noted that for all practical purposes, a compression of the sheet by a factor of $2$ or so would not affect our results qualitatively, since, again, the case $H/z_{ref}\ll 1$
would be reached earlier, thus just simplifying the arguments presented above.

\section{Gravitationally Unstable Modes in Sheets}\label{a:sheet}
The frequency and wavenumber for the gravitationally most unstable mode in an infinite, magnetized sheet were derived from equations~13, 29, and 30 given
by \citet{1985MNRAS.214..379L}. Specifically, we numerically found the maximum of the disperson relation $\Omega(\nu,\mu)$, and fit a 5th-order polynomial of the form
\begin{equation}
  \Omega_{max}(\mu) = \sum_0^5c_i \mu^i
\end{equation}
to the values $\Omega(\mu)$ and $\nu(\mu)$. Here, $\Omega=\imath \omega H/c_s$, and $\nu=k H$. The strength of the magnetic field is set by the magnetic
criticality parameter 
\begin{equation}
  \mu = \frac{2\pi\sqrt{G}\Sigma}{B}.
\end{equation}
The actual coefficients are listed in Table~\ref{t:omaxpol}. The fits are accurate to $<2$\% between $1.1<\mu<18$, and $<10$\% for $1.009<\mu<30.1$. Beyond
those values of $\mu$, the fits should not be used.

\begin{table*}
\begin{center}
\caption[Fit Parameters for $\Omega_{max}$ and $\nu_{max}$]{Fits are limited to $1.009<\mu<30.1$.\label{t:omaxpol}}
\begin{tabular}{lcccccc}
\hline
              & $c_0$        & $c_1$         & $c_2$     & $c_3$       & $c_4$       & $c_5$         \\
\hline
$\Omega_{max}$ &$0.0027172332$&$-0.093767163$&$1.1779378$&$-1.1181161$ &$0.38624833$ &$-0.043026629$ \\
$\nu_{max}$    &$0.0029897751$&$-0.089183578$&$1.0381798$&$-0.69517753$&$0.074091029$&$0.027382362$  \\
\hline
\end{tabular}
\end{center}
\end{table*}

\section{Diffusion Due to Unbending of Fieldlines}\label{a:horizontal}
If horizontal compressions occur in an accreting sheet, the field lines will bend slightly. This leads to a restoring force on the order of $\sim B^2(1+\tan \alpha)/(2\pi H)$, with
$\tan\alpha\ll 1$ as the angle between the horizontal field perturbation and the vertical background field, and $H$ the scale height of the sheet. The restoring force can lead to 
two effects. If the mass loading of the field lines is low (i.e. the inertia of the sheet is small), the field lines can straighten without diffusion, i.e. $\Sigma/B$ is conserved.
This would be essentially a magnetosonic wave in the sheet plane. If the mass loading is high, the field might unbend by diffusing horizontally through the sheet. The timescale would 
be $H^2 \tan^2\alpha/\lambda_{AD}$, i.e. equation~\ref{e:tauADapp} multiplied by $\tan^2\alpha$. For small angles, this could reduce the AD timescale substantially. However, for 
a one-dimensional, horizontal compression (assuming symmetry), $\Sigma/B$ would increase only by a factor of $(1+2\tan\alpha)$, so that this mechanism would play a role only 
when $\mu\approx 1$ already. The situation of the sheet as discussed here is different from that in accretion disks \citep[e.g.][]{1994MNRAS.267..235L,2014ApJ...785..127O}, where
a steady-state solution can be found between (radial) field advection and diffusion.

\bibliographystyle{mn2e}
\bibliography{./references}

\label{lastpage}

\end{document}